# A QoS-Aware Routing Protocol for Real-time Applications in Wireless Sensor Networks

Mary Cherian (Corresponding author)

Dr.Ambedkar Institute Of Technology, Outer Ring Road, Mallathahalli

Near Jnanabharathi Campus , Bangalore-560056, Karnataka, India

E-mail: thamasha2005@yahoo.com

T.R Gopalakrishnan Nair

Vice President, Research and Industry centre, DSC Institutions, Bangalore, Karnataka, India

**Abstract**

The paper presents a quality of service aware routing protocol which provides low latency for high priority packets. Packets are differentiated based on their priority by applying queuing theory. Low priority packets are transferred through less energy paths. The sensor nodes interact with the pivot nodes which in turn communicate with the sink node. This protocol can be applied in monitoring context aware physical environments for critical applications.

**Keywords**: Wireless Sensor Network (WSN), Quality of service, Pivot Node, Queuing model.

## 1. Introduction

A wireless sensor network is a wireless network consisting of spatially distributed autonomous devices using sensors to cooperatively monitor physical or environmental conditions, such as temperature, sound, vibration, pressure, motion or pollutants, at different locations. The development of wireless sensor networks was originally motivated by military applications such as battlefield surveillance. However, these networks are now used in many industrial and civilian application areas, including industrial process monitoring and control, machine health monitoring, environment and habitat monitoring, healthcare applications, home automation, and traffic control. In addition to one or more sensors, each node in a sensor network is typically equipped with a radio transceiver or other wireless communication device, processor and battery (Akyildiz *et al.* 2002).The environment under surveillance can be populated with sensors for capturing the context information and communicating to the base station for further action (Intanagonwiwat *et al.* 2000).

### 1.1 QoS aware routing protocols

The main goal of QoS aware routing protocols is to find a path to meet the QoS requirements. The routing protocol searches for lower cost and energy efficient path between end-to end connections (Heinzelman *et* *al.* 2000). This paper introduces the Pivot node which is different from the sensor nodes. Pivot node has better processing and communication power, and longer lifetime batteries. Pivot nodes are added between the sensor nodes and the sink node.

The network QoS aims at defining the bandwidth, latency, jitter that each packet expects from the network. As the bandwidth in each link in a network is fixed, the bandwidth given for each flow depends on the QoS requirement of the particular flow. A QoS architecture specifies the performance requirements for different types of packets and methods for delivering those performance requirements (Carpenter *et al.* 2002). The network and applications use three types of communication schemes. They are periodic interactions, interactions triggered by certain events and interactions through queries (Boukerche *et al.* 2004). Such kinds of communication schemes have some requirements. These requirements include: low latency delivery of high priority packets, packet delivery reliability, fast path repair in the presence of failures, uniform energy saving and support to differentiated service packets.

The rest of the paper is organized as follows: Section 2 discusses the related work. Section 3, gives the protocol description. Section 4, indicates simulation and results. Section 5, draws conclusions and provides the future works possible.

## 2. Related Work

In this section some background studies and the work done for QoS for WSN are explained.

The energy aware QoS routing protocol developed for Wireless Sensor Networks (Akkaya *et al.* 2003), considers energy efficiency to maximize the network lifetime. A least-cost, delay-constrained path is found for real-time data in terms of link cost that captures nodes energy reserve, transmission energy, error rate and other communication parameters. But this protocol does not consider the imprecise state information while determining the routes. A survey on routing protocols for wireless sensor networks (Akkaya *et al.* 2005), details the recent routing protocols for sensor networks and presents a classification for the various approaches pursued. The three main sensor networks described in this paper are data-centric, hierarchical and location-based. Protocols using contemporary methodologies such as network flow and quality of service modeling are also discussed. This survey does not include information about spatial queries and databases using distributed sensor nodes and the interaction with the location-based routing protocols. Sequential Assignment Routing (SAR) (Sohrabi *et al.* 2002) is one of the QoS aware routing protocols applied to plane networks. SAR uses tables and multiple paths to save energy and achieve fault tolerance. SAR protocol creates trees of nodes considering the residual energy in each path as well as the priority level of each packet. Simulation results show that SAR can provide low energy dissipation. SAR can keep multiple paths from the nodes to the sink what guarantees fault tolerance. But it suffers from high overhead to keep the routing tables and states at each node. The overhead increases for dense networks. Priority based Congestion Control Protocol (C. Wang *et al.* 2007) proposes a node priority based control mechanism for wireless sensor networks. But PCCP has limitation for handling multiple sensed data at a node.

## 3. Protocol description

This paper focuses on plane and hierarchical networks.   It presents QoS- Aware Routing Protocol and also performs an analysis of the proposed protocol in meeting the requirements of QoS for the Emergency Class of Applications. The proposed protocol is designed with certain considerations about the network model and the queuing model. Figure 1 depicts the network model with sensor nodes as Si with sensor nodes S1 to S9 delivering data to the pivot node. The pivot node in turn communicates with the sink node. The lines between nodes indicate the path of communication.

The important criteria considered for this network model are

-Each node can sense heterogeneous application data.
-Pivot node should work as intermediate sink for sensor nodes.
-Pivot node should have more communication capability than sensor nodes.

-All the nodes have the same transmission range.

The queuing model in Figure 2 is implemented at every sensor node, to differentiate between high priority packets and low priority packets. The queuing model is specifically designed for the coexistence of real-time and non-real time traffic in each sensor node (Bhuyan, *et al.* 2010). The model we employ is inspired from class-based queuing model (Akkaya *et al.* 2003). Packets are labeled according to their type. We use different queues for the two different types of traffic. We assume that the MAC layer is collision free and will not create significant delay.

*3.1 Protocol Description*

The configuration of the network takes place through the construction of hop trees.

Initially, hop1 tree construction is carried out by sink node, which initializes the attribute SHOT1 with value 0. The SHOT1 value 0 indicates the energy level at the sink node. Each sensor node knows only the information about its neighbors. Each sensor node receives SHOT1 value, saves it, increments it and sends to all the neighbors. It uses a routing technique called flooding. Loop back is avoided by comparing received SHOT1 value with saved SHOT1 value. Lower SHOT1 value is accepted. The pivot node also starts constructing hop2 tree. It initializes attribute SHOT2 value with the energy level 0. At the end of hop tree construction, each sensor node consists two energy levels (a,b) where a indicates number of hops from the sink node and b indicates number of hops from the pivot node. After the hop tree construction, subscription of messages takes place. The sink node informs its interest to sensor nodes, each sensor node consists of a requirement table and a routing table. The interest is stored in the requirement table by each node.  When sensor node captures some event, it compares it with the entry of the requirement table. Accordingly finds best path from routing table to forward the report about the event to the sink of that interest. While sending report message, the sensor node which captures event sends packet to pivot node by hop-by-hop, and then the pivot node sends it directly to the sink node. The path used to send subscription message from the sink node to the sensor node is used in reverse to send report by the sensor node to the sink node in the same area. The packets are differentiated using Fair Queuing (FQ) technique (Peterson 2005, pp.463-465). The organizer in the queuing model sends the real time data packets which has higher QoS requirement to the real-time queue, while the non-real time packets are sent to the second queue.

Packets having highest priority are scheduled first by the scheduler; it provides low latency for higher priority packets. The priority of the packets is assigned by the two bits priority field in the header of the packet. The field will have values 0,1,2,3 with 3 being the highest priority.

Path updating mechanism is also incorporated in this protocol. It is an ACK-based mechanism. The sender node sends a packet and starts a timer, waiting for an ACK message. If it does not receive an ACK message within the time out period the node selects an alternate path from the routing table. It sends HELLO message to all its neighbors, which responds with a REPLY message consisting its hop level and ID. The neighbor with the least hop level is selected as the new receiver.

## 3  Simulation Results and Discussions

The simulation determines the latency of real time packets and gives a comparison between the latency of real time packets and non-real time packets. It also indicates the variation in latency with the increase in network size.

The details of simulation parameters are as follows: In an area of 50mx50m sensor field 50 sensors are deployed randomly. Sensors are having a transmission range of 10 m and pivot nodes are having a transmission range of 100 m. Pivot nodes are provided with more transmission capabilities than sensor nodes. It is assumed that there is no interference from other nodes.

The lifetime of a wireless sensor network is constrained by the limited energy and processing capabilities of its nodes. Sensor nodes have limited energy. To extend the life time of the sensor networks it is very important to have high energy efficiency at all the processing nodes (Cherian *et al.* 2011). As the pivot nodes collect the data from the sensor nodes, which in turn forward the data to the sink node,the energy consumption is less since the number of hops to reach the sink node from the sensor node is less.

The diagrams below illustrate the different phases of the protocol, hop tree construction, propagation of the subscription messages, and the data collection by the pivot node and the sink node.

Figure 3 depicts the status of sensor node 4 after the HOP tree construction by the sink node.
Figure 4 indicates the subscription of the interest by the sink node. Here the sink requires the temperature of the specific area to be gathered and delivered by the sensor nodes monitoring the specific area.
Figure 5 indicates pivot node receiving the report message from the sensor nodes about the interest. Figure 6 depicts the propagation of the collected data by the pivot node to the sink node.
The pivot node promotes the interaction between sink node and the sensor node. Figure 7 illustrates the relationship between the network size and the average latency. It indicates that latency for real time and non real time packets increases with the increasing network size. The protocol provides low latency for real time packets when network size increases. Network Size indicates the number of nodes in the entire network which consists of sink nodes, pivot nodes and the sensor nodes. The protocol is capable of keeping a high delivery rate for real time packets, even in the presence of path failures.

## 5   Conclusion

Wireless networks consisting of pivot nodes and sensor nodes monitor the physical environment for real time data collection. Energy consumption is less, since the number of hops required to reach the sink node is less, since the data is collected by the pivot node, and forwarded to the sink node. The queuing model supports the delivery of high priority real time packets at a higher rate, minimizing the end to end delay.

We have not included a battery model in this work to account for the energy constrains in the wireless sensor networks. The protocol assumes an ideal MAC layer and the future work can include the radio media access at the MAC layer. The work can be extended to other topologies. We can also use a predictive QOS model. Integration of sensor networks with IP based Internet can also be considered.
To deal with the energy constraints of the sensor node a sleep mode for each node may be included where the sensor node that has not been used for long goes to sleep state and may wake upon the reception of the sensed data.

**Acknowledgment**

Authors thank Bilwa R Sarode of Dr.Ambedkar Institute of Technology, Bangalore for the support for the required simulation in the laboratory.

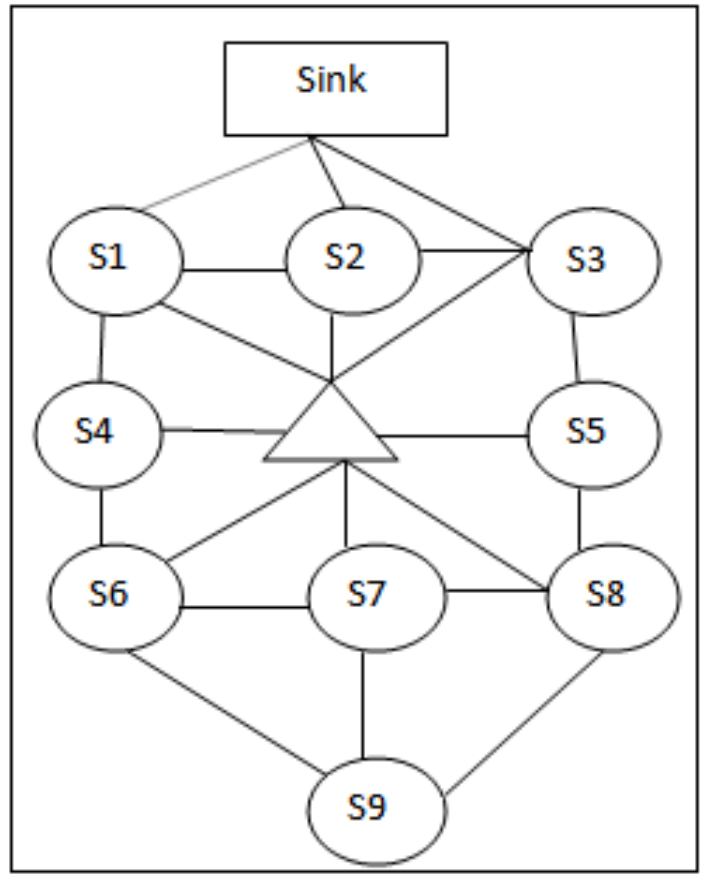

Figure 1    Network Model

Figure 1 depicts the network model with sensor nodes S1 to S9 delivering data to the pivot node. Pivot node is indicated by a triangle. The lines between nodes indicate the path of communication.

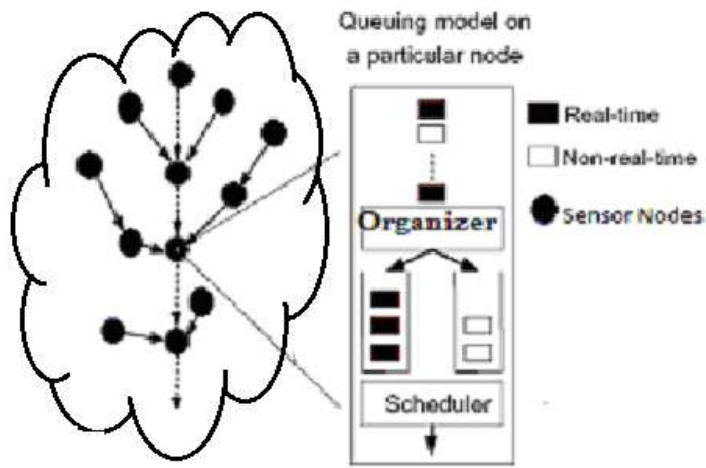

Figure 2     Queuing Model

The queuing model in figure 2 is implemented at every sensor node, which differentiates between high priority packets and low priority packets. One queue is assigned for real time packets with higher priority and another one is assigned for non real-time packets with lower priority.

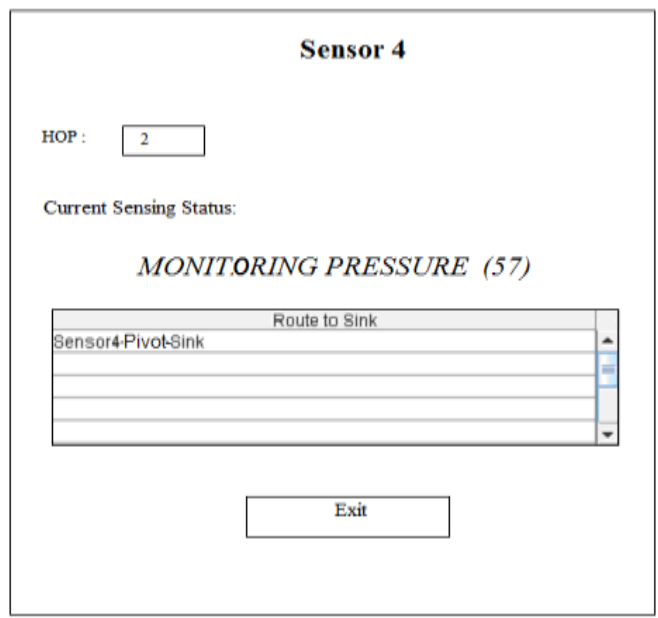

Figure 3    HOP tree constructions

Figure 3 depicts the status of sensor node 4 after the HOP tree construction by the sink node.

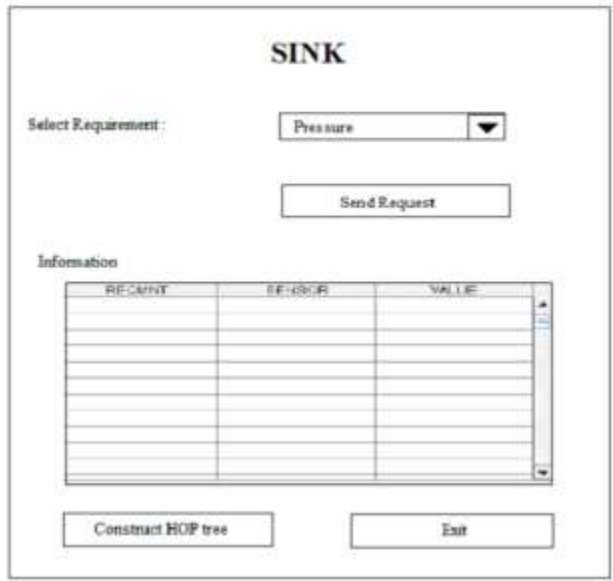

Figure 4   Sink window propagation of subscription message, which indicates the subscription of the interest by the sink node.

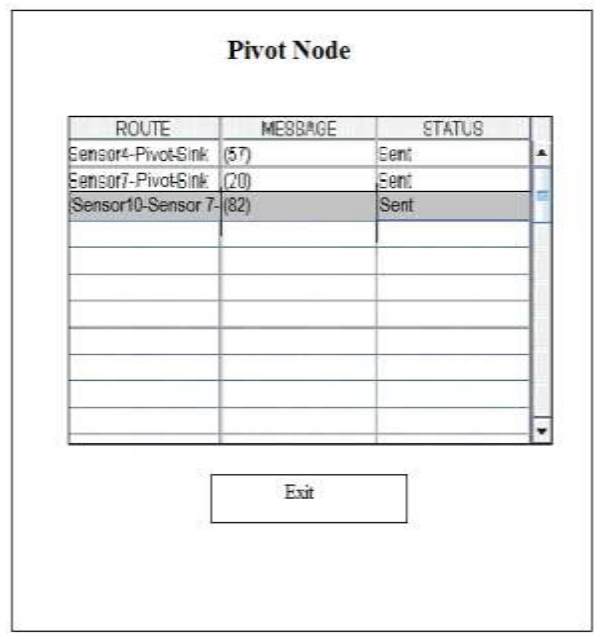

Figure 5   Pivot node receiving data from the sensor nodes, which indicates pivot node receiving the report message from the sensor nodes about the interest.

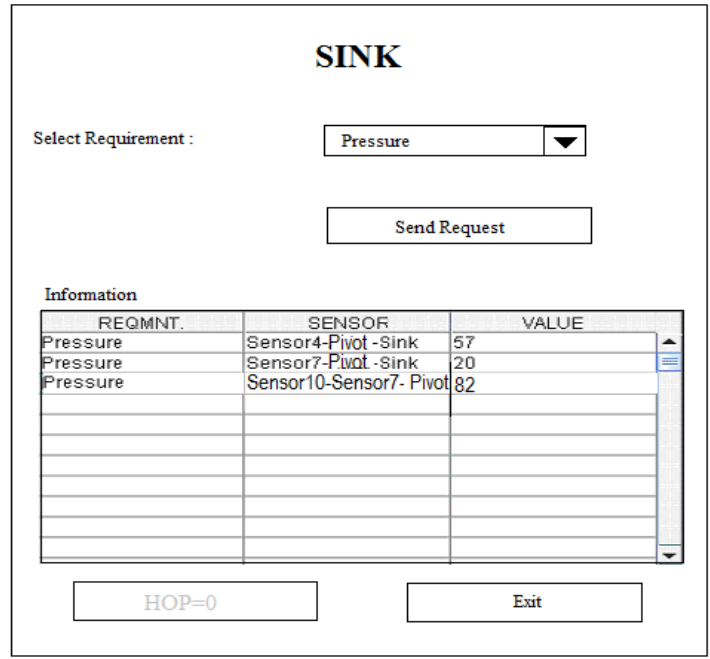

Figure 6    Report message from pivot to the sink node, which depicts the propagation of the collected data by the pivot node to the sink node.

Table 1.    Network size verses Latency

| Network size | Average Latency (Real Time Packets) | Average Latency (Non Real Time Packets) |
|---|---|---|
| 10 | 0.12 | 0.2 |
| 15 | 0.2 | 0.4 |
| 20 | 0.31 | 0.5 |
| 25 | 0.4 | 0.7 |

The table gives the average latency for real time and non-real time packets for different network sizes.

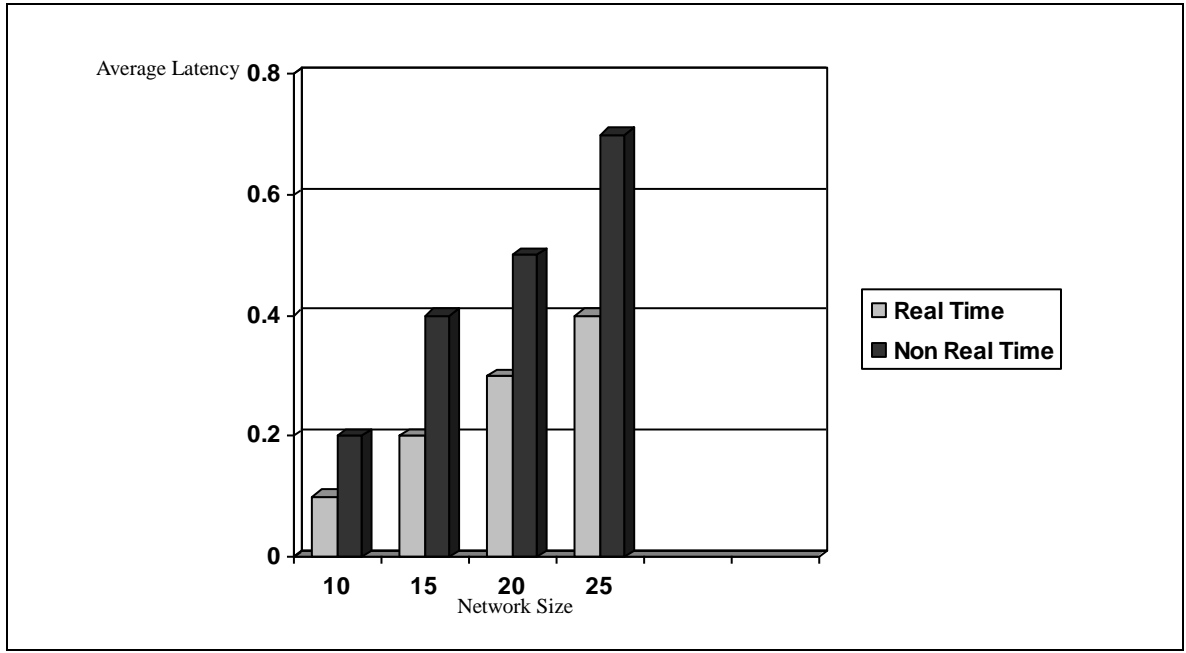

Figure 7    Network   Size verses   Average Latency

Figure 7 illustrates the relationship between the network size and the average latency. The protocol provides low latency for real time packets when the network size increases